\documentclass[manuscript]{aastex}
\usepackage{emulateapj5}
\usepackage{apjfonts}
\usepackage{times}
\usepackage{natbib}
\makeatletter
\newenvironment{inlinetable}{%
\def\@captype{table}%
\noindent\begin{minipage}{0.999\linewidth}\begin{center}\footnotesize}
{\end{center}\end{minipage}\smallskip}

\newenvironment{inlinefigure}{%
\def\@captype{figure}%
\noindent\begin{minipage}{0.999\linewidth}\begin{center}}
{\end{center}\end{minipage}\smallskip}
\makeatother

\def\ghz{\rm GHz}
\def\lsim{\mathrel{\rlap{\lower4pt\hbox{\hskip1pt$\sim$}}
    \raise1pt\hbox{$<$}}}                
\def\gsim{\mathrel{\rlap{\lower4pt\hbox{\hskip1pt$\sim$}}
    \raise1pt\hbox{$>$}}}                

\usepackage{graphicx}
\shorttitle{ Angular power spectrum of  WMAP CMB anisotropy  }

\begin{document}
\bibliographystyle{apjl} 

\title{A blind estimation of the angular power spectrum of CMB anisotropy from WMAP} 

 \author{Rajib Saha\altaffilmark{1,2}, Pankaj Jain\altaffilmark{1},
 Tarun Souradeep\altaffilmark{2}}

 \altaffiltext{1}{Physics Department, Indian Institute of Technology,
 Kanpur, U.P, 208016, India.} 
 \altaffiltext{2}{IUCAA, Post Bag 4, Ganeshkhind, Pune 411007, India. }

\begin{abstract}

Accurate measurements of angular power spectrum of Cosmic Microwave
Background (CMB) radiation has lead to marked improvement in the
estimates of different cosmological parameters. This has required
removal of foreground contamination as well as detector noise bias
with reliability and precision.  We present the estimation of CMB
angular power spectrum from the multi-frequency observations of WMAP
using a novel {\em model-independent} method. The primary product of
WMAP are the observations of CMB in 10 independent difference
assemblies (DA) that have uncorrelated noise.  Our method utilizes
maximum information available within WMAP data by linearly combining
all the DA maps to remove foregrounds and estimating the power
spectrum from cross power spectra of clean maps with independent
noise.  We compute $24$ cross power spectra which are the basis of the
final power spectrum. The binned average power matches with WMAP
team's published power spectrum closely. A small systematic difference
at large multipoles is accounted for by the correction for the
expected residual power from unresolved point sources. The correction
is small and significantly tempered. Previous estimates have depended
on foreground templates built using extraneous observational input.
This is the {\em first demonstration that the CMB angular spectrum can
be reliably estimated with precision from a self contained analysis of
the WMAP data}.
\end{abstract}
\section{Introduction}

Remarkable progress in cosmology has been made due to the measurements
of the anisotropy in the cosmic microwave background (CMB) over the
past decade.  The extraction of the angular power spectrum of the CMB
anisotropy is complicated by foreground emission within our galaxy and
extragalactic radio sources, as well, as the detector
noise~\cite{bouc_gisp99,Tegmark96}. It is established that the CMB
follows a blackbody distribution to high accuracy, \cite{FIRAS}.
Hence, foreground emissions may be removed by exploiting the fact that
their contributions in different spectral bands are considerably
different while the CMB power spectrum is same in all the
bands~\cite{Dodelson, Tegmark2000a, Bennett1,Tegmark98}.  Different
approaches to foreground removal have been proposed in the
literature~\cite{bouc_gisp99,Tegmark96,hob98} \\ ~\cite{main0203,erik05}

The Wilkinson Microwave Anisotropy Probe (WMAP) observes in $5$
frequency bands at $23~\ghz$~(K), $33~\ghz$~(Ka), $41~\ghz$~(Q),
$61~\ghz$~(V) and $94~\ghz$~(W).  In the first data release, the WMAP
team removed the galactic foreground signal using a template fitting
method based on a model of synchrotron, free free and dust emission in
our galaxy \cite{Bennett1}. The sky map around the galactic plane and
around known extragalactic point sources were masked out and the CMB
power spectrum was then obtained from cross power spectra of
independent difference assemblies in the $41~\ghz$, $61~\ghz$ and
$94~\ghz$ foreground cleaned maps~\cite{Hinshaw}.

A model independent removal of foregrounds has been proposed in the
literature \\ ~\cite{Tegmark96}.  The 
method has also
been implemented on the WMAP data in order to create a foreground
cleaned map~\cite{Tegmark}. The main advantage of this method is that
it does not make any additional assumptions regarding the nature of the
foregrounds.  Furthermore, the procedure is computationally fast.  The
foreground emissions are removed by combining the five different WMAP
bands by weights which depend both on the angular scale and on the
location in the sky (divided into regions based on `cleanliness').
However, this analysis 
did not
attempt to remove the detector noise
bias~\cite{Tegmark}. Consequently, the power spectrum recovered from
the foreground cleaned map has a lot of excess power at large multipole
moments due to amplification of detector noise bias beyond the beam
resolution.

The prime objective of our paper is to remove detector noise bias
exploiting the fact that it is uncorrelated among the different
Difference Assemblies (DA) \cite{Hinshaw,Jarosik}. The WMAP data uses
10 DA's \cite{Bennett, Bennett2, WMAP, Hinshaw1}, one each for K and
Ka bands, two for Q band, two for V band and four for W band.  We
label these as K, Ka, Q1, Q2,V1, V2, W1, W2, W3, and W4
respectively. We eliminate the detector noise bias using cross power
spectra and provide a model independent extraction of CMB power
spectrum from WMAP first year data.  So far, only the $3$ highest
frequency channels observed by WMAP have been used to extract CMB
power spectrum and the foreground removal has used foreground
templates based on extrapolated flux from measurements at frequencies
far removed from observational frequencies of
WMAP~\cite{Hinshaw,Pablo,Patanchon}.  We present a more general
procedure where we use observations from all the $5$ frequency
channels of WMAP and do not use any extraneous observational input.

\section{Methodology}
\subsection{Foreground Cleaning}

Up to the foreground cleaning stage, our method is similar to Tegmark
\& Efstathiou (1996) and Tegmark {\it et al.} (2003). In Tegmark {\it
et al.} (2003), a foreground cleaned map is obtained by linearly
combining 5 maps corresponding to one each for the different WMAP
frequency channels. For the Q, V and W frequency channels, where more
than one maps were available, an averaged map was used. However,
averaging over the DA maps in a given frequency channel precludes any
possibility of removing detector noise bias using cross
correlation. In our method we linearly combine maps corresponding to a
set of 4 DA maps at different frequencies. We treat K and Ka maps
effectively as the observation of CMB in two different DA. Therefore
we use K and Ka maps in separate combinations. In case of W band 4 DA
maps are available. We simply form an averaged map taking two of them
at a time and form effectively 6 DA maps. 
 W$ij$ represents simply an averaged map obtained from the $i^{\rm th}$ and
$j^{\rm th}$ DA of W band. (Other variations are possible. We defer a
discussion to a more detailed publication~\cite{Saha}).
In table \ref{tab:combination} we list all the $48$ possible linear combinations
of the DA maps that lead to `cleaned' maps, C${\bf i}$ and CA${\bf i}$'s, where
{${\bf i}$ = 1, 2, \ldots, 24}.

\begin{inlinetable}
\scriptsize
\begin{tabular}{|l |l|}
\hline 
                         &                           \\
(K,KA)+Q1+V1+W12=(C1,CA1)&(K,KA)+Q1+V2+W12=(C13,CA13)\\
(K,KA)+Q1+V1+W13=(C2,CA2)&(K,KA)+Q1+V2+W13=(C14,CA14)\\
(K,KA)+Q1+V1+W14=(C3,CA3)&(K,KA)+Q1+V2+W14=(C15,CA15)\\
(K,KA)+Q1+V1+W23=(C4,CA4)&(K,KA)+Q1+V2+W23=(C16,CA16)\\
(K,KA)+Q1+V1+W24=(C5,CA5)&(K,KA)+Q1+V2+W24=(C17,CA17)\\
(K,KA)+Q1+V1+W34=(C6,CA6)&(K,KA)+Q1+V2+W34=(C18,CA18)\\
(K,KA)+Q2+V2+W12=(C7,CA7)&(K,KA)+Q2+V1+W12=(C19,CA19)\\
(K,KA)+Q2+V2+W13=(C8,CA8)&(K,KA)+Q2+V1+W13=(C20,CA20)\\
(K,KA)+Q2+V2+W14=(C9,CA9)&(K,KA)+Q2+V1+W14=(C21,CA21)\\ 
(K,KA)+Q2+V2+W23=(C10,CA10)&(K,KA)+Q2+V1+W23=(C22,CA22)\\ 
(K,KA)+Q2+V2+W24=(C11,CA11)&(K,KA)+Q2+V1+W24=(C23,CA23)\\ 
(K,KA)+Q2+V2+W34=(C12,CA12)&(K,KA)+Q2+V1+W34=(C24,CA24)\\
                           &                           \\
 \hline 
\end{tabular} 
 \caption{ 
Different combinations of the DA maps used to
 obtain the final 48 cleaned maps.}
\label{tab:combination}
\end{inlinetable}



Following the approach of Tegmark \& Efstathiou (1996), we introduce a
set of weights, ${{\bf [W_l]}=(w_l^1, w_l^2, w_l^3, w_l^4)}$ for each
of the $4$ DA in the combination,
which defines our cleaned map as the 
linear combination
\begin{equation}
a_{lm}^{\rm Clean}=\sum_{i=1}^{i=4}w_l^{i}\frac{a_{lm}^i}{B_l^i}\,,
\label{c_map}
\end{equation}
where $a_{lm}^i$ is spherical harmonic transform of map and $B^i_l$ is
the beam function for the channel $i$ supplied by the WMAP team.

\noindent The condition that the CMB signal remains untouched during
cleaning is encoded as the constraint $\bf [W_l][e]=[e]^T[W]^T=1$
where  ${\bf [e] }$ is a $4\times 1$ column vector with unit elements.


Following Tegmark \& Efstathiou (1996), Tegmark {\it et al.} (2003)
and Tegmark {\em et al.} (2000a) we obtain the optimum weights which
combine $4$ different frequency channels subject to the constraint
that CMB is untouched,
$\bf [W_l]={[e]^T[C_l]^{-1}}/\left({[e]^T[C_l]^{-1}[e]}\right)$.
Here the matrix ${\bf [C_l]}$ is 
\begin{equation}
{\bf [C_l]} \equiv C^{ij}_l=\frac{1}{2 l+1}\sum_{m=-l}^{m=l}
\frac{a^{i}_{lm}a^{j*}_{lm}}{B_l^i B_l^j}\,.
\end{equation} 

In practice, we smooth all the elements of the ${\mathbf C_l}$ using a moving average window over $\Delta l = 11$ before deconvolving by the beam function. 
This avoids the possibility of an ocassional singular ${\mathbf C_l}$ matrix.
The entire cleaning procedure is automated and takes
approximately $3$ hours on a $16$ alpha processor machine to get
the $48$ cleaned maps. One of the cleaned maps, C8, is shown in the
Fig.~\ref{c_map1}. In all the $48$ maps some residual foreground
contamination is visibly present along a small narrow strip on the
galactic plane.  In a future publication, we assess the quality of
foreground cleaning in these maps using the Bipolar power spectrum
method~\cite{Amir1,Amir2,Amir3} and compare them to others such as the
internal Linear combination map (ILC) of WMAP.  For the angular power
estimation that follows, the Kp2 mask employed suffices to mask the
contaminated region in all the $48$ maps.

\begin{inlinefigure}
\centerline{\includegraphics[scale=0.35,angle=90]{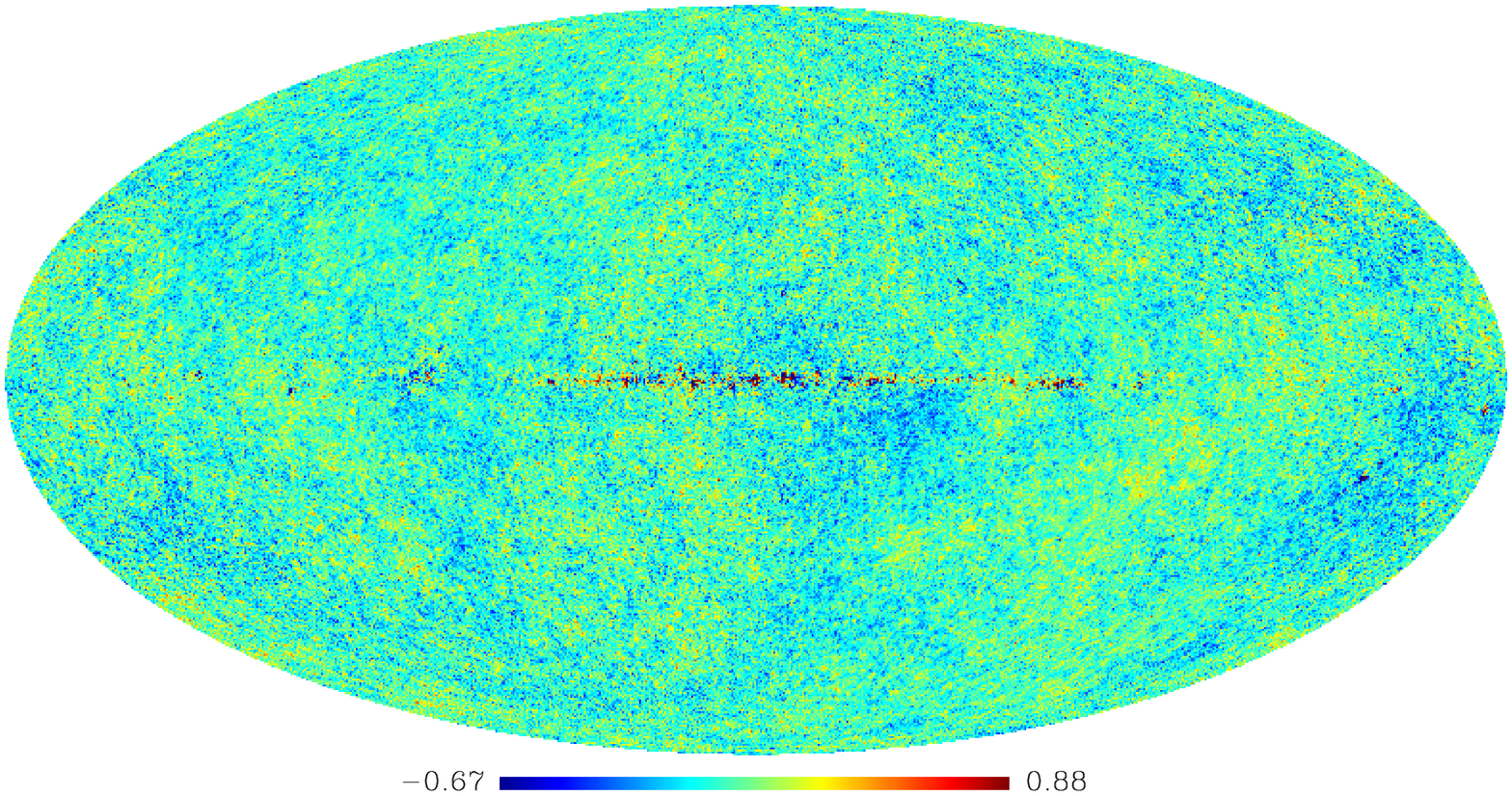}}
\caption { The cleaned map $C8$ for the K, Q2, V2, W13 combination
(units ${\rm mK}$). Residual foreground contamination visible along
the galactic plane are well within the Kp2 mask that is applied before
extracting the angular power spectrum.}
\label{c_map1}
\end{inlinefigure}

\subsection{Power Spectrum Estimation}
\label{power_spectrum_estimation}

We obtain cross correlated power spectrum from these cleaned maps
after applying a Kp2 mask. In choosing pairs C{\bf i} \& C{\bf j} to be
cross correlated, we ensure that no DA is common between them.
 Figure \ref{each_bin} lists and plots
the $24$ cross power spectra for which the noise bias is zero.  The
cross power spectra are corrected for effect of the mask, the beam and
pixel window. These are accounted for by de-biasing the pseudo-$C_l$
estimate using the coupling (bias) matrix corresponding to the Kp2
mask and appropriate circularized beam transform~\cite{Hivon}.  Figure
\ref{each_bin} plots the $24$ cross power spectra (binned)
individually. The spectra closely match each other for $l< 540$. The
$24$ cross power spectra are then combined with equal weights into a
single `Uniform average' power spectrum ~\footnote{There exists the
additional freedom to choose optimal weights for combining the 24
cross-power spectra which we do not discuss in this work.}.  We also
estimate the residual power contamination in the `Uniform average'
power spectrum from the unresolved point sources on running through
our analysis the same source model used by WMAP team to correct for
this contaminant~\cite{Hinshaw}. The model is derived entirely within
the WMAP data based on fluxes and spectra of $208$ resolved point
sources identified in the maps~\cite{Bennett1}.  The residual power
from unresolved point sources is a constant offset of $\sim 140\mu
K^2$ for $l \gsim 400$ (and negligible at lower $l$).
This residual is much less than actual point source contamination in
Q, KA or K band and intermediate between V and W band point source
contamination. It is noteworthy that the method significantly tempers
the point source residual at large $l$ that otherwise  is $\propto l^2$
in each map.  The final power spectrum is binned in the same manner as
the WMAP's published result for ease of comparison.

\subsection{Error Estimate on the Power Spectrum}
\label{Errorbars}
The errors on the final power spectrum are computed from $110$ random
Monte Carlo simulations of CMB maps for every DA each with a
realization based on corresponding WMAP noise map (available at
LAMBDA) and diffuse foreground contamination. The common CMB signal in
all the maps was based on a realization of the WMAP `power law' best
fit $\Lambda$-CDM model~\cite{Spergel}. We use the publicly available
Planck Sky Model to simulate the contamination from the diffuse
galactic (synchrotron, thermal dust and free-free) emission at the
WMAP frequencies. The CMB maps were smoothed by the beam function
appropriate for each WMAP's detector.  The set of DA maps
corresponding to each realizations were passed through the same
pipeline used for the real data. Averaging over the $110$ power
spectra we recover the model power spectrum, 
but for a hint of bias towards lower values in the low $l$ moments. For $l=2$ and $l=3$ the bias is $-27.4\%$ and $-13.8\%$ respectively. However this bias become negligible at higher $l$, e.g. at $l=22$, it is only $-0.8\%$.


The standard deviation obtained from the diagonal elements of the
covariance matrix is used as the error bars on the $C_l$'s obtained
from the data. (The beam uncertainty is not included here, but is
deferred to future work where we also incorporate non-circular beam
corrections~\cite{Sanjit}.) 

\begin{inlinefigure}
\includegraphics[scale=0.35,angle=-90]{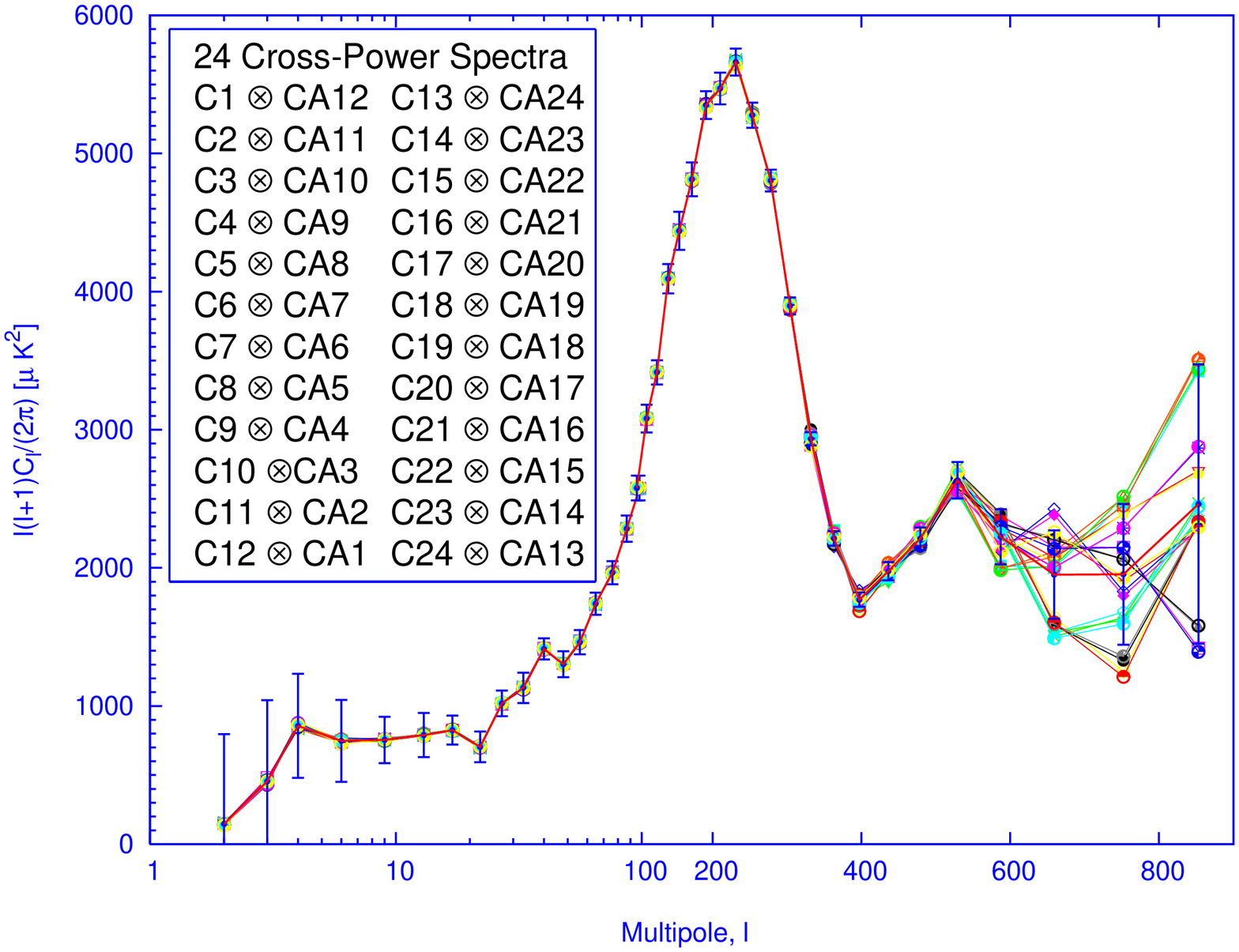}
\caption{The $24$ individual cross power spectra corresponding to the
cross correlations listed in this figure are plotted. The cross spectra
show very small dispersion for $l\lsim 540$. The average power
spectrum is plotted in red line and blue error bars. The multipole range
is on log scale for $l<100$, and linear, thereafter.}
\label{each_bin}
\end{inlinefigure}

\section{Results} 
\label{res1}
We obtain a `Uniform average' power spectrum of the 24 cross spectrum
following the method mentioned in
section~\ref{power_spectrum_estimation}.  
The black curve in Fig.~\ref{Cl_uniform_weight} compares our results
with the WMAP published power spectrum plotted with red line. The
power spectra from two independent analysis are reasonably close.  In
case of `Uniform average' a maximum difference of $92~\mu K^2$ is
observed only for octopole. For the large multipole range the
difference is small and for $l=752$ it is approximately $48~\mu
K^2$. This is well within the $1\sigma$ error bar ($510 \mu K^2$ )
obtained from the simulation.  The small, but systematic excess, at
large multipoles is precisely resolved when our `Uniform average' is
corrected for the expected residual power from unresolved point source
contamination described in $\S$\ref{power_spectrum_estimation}. The
point source corrected power spectrum is shown in black line in this
figure. The difference of this power spectrum with the published WMAP
estimate is shown in the bottom panel of
Fig.~\ref{Cl_uniform_weight}. The differences are well within the
$1\sigma$ error-bars estimated from the 
simulations described in $\S$\ref{Errorbars}.

\begin{inlinefigure}
\includegraphics[scale=0.35,angle = -90]{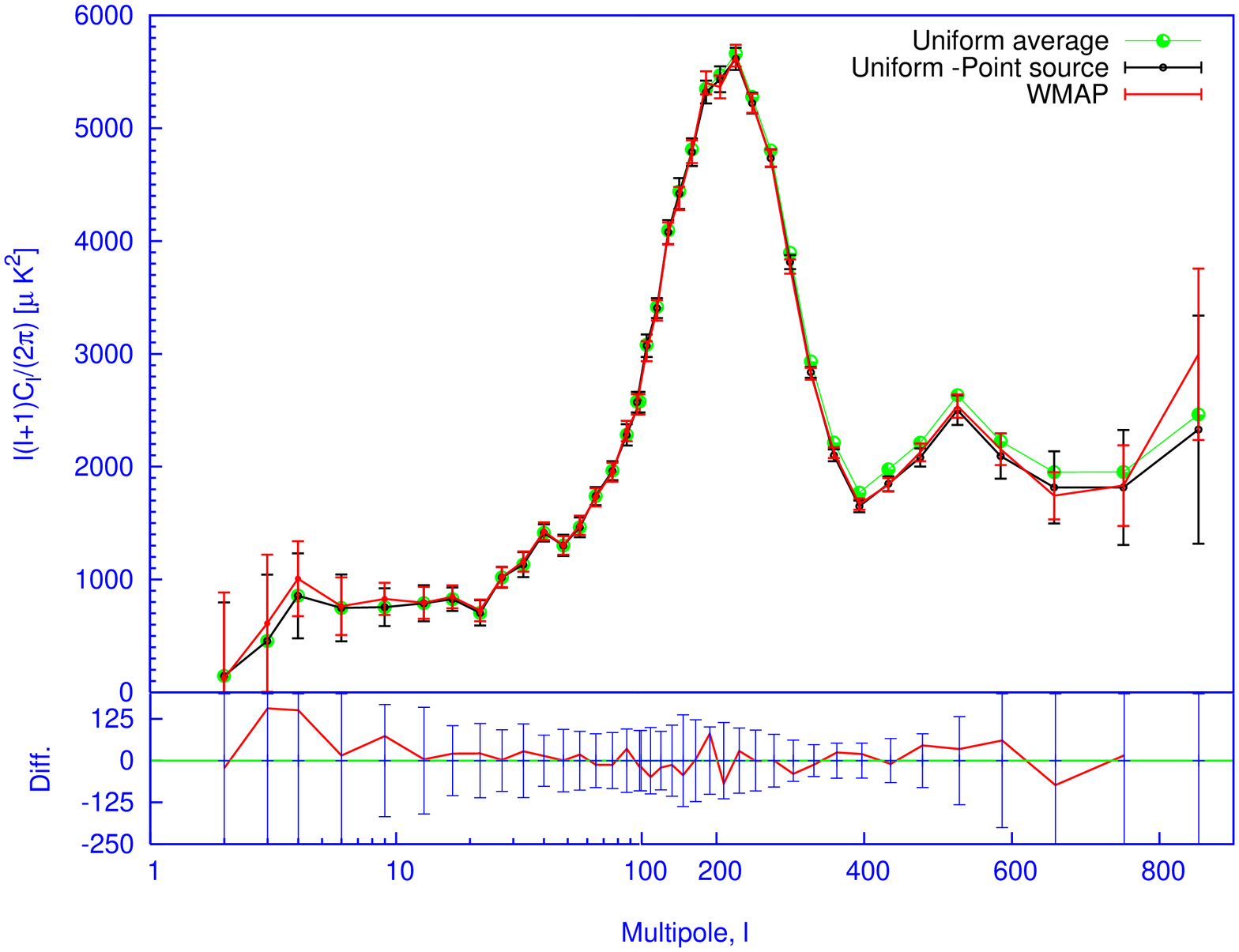}
\caption{ 
The `Uniform average power spectrum is plotted in green line. The
black line shows residual unresolved point source corrected power
spectrum. The error-bars are computed from the diagonal elements of the
covariance matrix obtained from our simulation pipeline.  Beyond $l
=400$ all the error bars are shifted by $\delta l = 10 $ to visually
distinguish between error bars obtained from our method and the WMAP's
published error bars. The published binned WMAP power spectrum is
plotted in red line with error bars.
The multipole range $2< l <100 $ are plotted in the log scale to
show the small $l$ behavior of the power spectrum. }
\label{Cl_uniform_weight}
\end{inlinefigure}

We find a suppression of power in the quadrupole and octopole moments
consistent with WMAP published result. However, our quadrupole moment
($ 146 \mu K^2 $) is a little larger than the WMAP's quadrupole moment
($ 123 \mu K^2 $) and Octopole ($ 455 \mu K^2$) is less than WMAP's
result ($ 611 \mu K^2$). 
The `Uniform average' power spectrum does not show the `bite' like
feature present in WMAP's power spectrum at the first acoustic peak
reported by WMAP~\cite{Hinshaw}. We perform a quadratic fit to the
peaks and troughs of the binned spectrum similar to WMAP
analysis~\cite{page}. For the residual point source corrected
(`Uniform average') power spectrum we obtain the first acoustic peak
at $l = 219.8 \pm 0.8 $ $(220.8 \pm 0.8 )$ with amplitude $\Delta T_l
= 74.1 \pm 0.3 \mu K $ $ (74.4 \pm 0.3 \mu K)$, the second acoustic
peak at $l = 544 \pm 17 $ $ (545 \pm 17 )$ with amplitude $\Delta T_l
= 48.3 \pm 1.2 \mu K $ $(49.6 \pm 1.2 \mu K)$ and the first trough at
$l = 419.2 \pm 5.6 $ $(418.7 \pm 5.5 )$ with amplitude $\Delta
T_l = 41.7 \pm 1 $ $ \mu K (42.2 \pm 0.9 \mu K)$.

As cross checks of the method, we have carried out analysis with other
possible combinations of the DA maps.

\begin{enumerate}
\item{} The WMAP team also provide foreground cleaned maps
corresponding to Q1 to W4 DA (LAMBDA). The Galactic foreground signal,
consisting of synchrotron, free-free, and dust emission, was removed
using the 3-band, 5-parameter template fitting
method~\cite{Bennett1}. We also include K and Ka band maps which are
not foreground cleaned. 
The resulting power spectrum from our analysis
matches closely to the `Uniform average' power spectrum.


\item{} Excluding the K and Ka band from our analysis we get a power
spectrum close to the `Uniform average' results. Notably, we find a
more prominent notch at $l = 4$ similar to WMAP's published results.
\end{enumerate}

This is a clear demonstration that the blind approach to foreground
cleaning is comparable in efficiency to that from template fitting
methods and certainly adequate for a reliable estimation of the
angular power spectrum. 

\section{Conclusion}

The rapid improvement in the sensitivity and resolution of the CMB
experiments has posed increasingly stringent requirements on the level
of separation and removal of the foreground contaminants.  Standard
approaches to foreground removal, usually incorporate the extra
information about the foregrounds available at other frequencies, the
spatial structure and distribution in constructing a foreground
template at the frequencies of the CMB measurements. These approach
could be susceptible to uncertainties and inadequacies of modeling
involved in extrapolating from the frequency of observation to CMB
observations.

We carry out an estimation of the CMB power spectrum from the WMAP
first year data that is independent of foreground model and evades
these uncertainties. The novelty is to make clean maps from the
difference assemblies and exploit the lack of noise correlation
between the independent channels to eliminate noise bias.  {\em This
is the first demonstration that the angular power spectrum of CMB
anisotropy can be reliably estimated with precision solely from the
WMAP data (difference assembly maps) without recourse to any external
data.}

The understanding of polarized foreground contamination in CMB
polarization maps is rather scarce. Hence modeling uncertainties could
dominate the systematics error budget of conventional foreground
cleaning.  The blind approach extended to estimating polarization
spectra after cleaning CMB polarization maps could prove to be
particularly advantageous.


\section{Acknowledgment}

The analysis pipeline as well as the entire simulation pipeline is
based on primitives from the Healpix package. {\footnote {The Healpix
distribution is publicly available from the website
http://www.eso.org/science/healpix.} } We acknowledge the use of
version 1.1 of the Planck reference sky model, prepared by the members
of Working Group 2 and available at www.planck.fr/heading79.html. The
entire analysis procedure was carried out on the IUCAA HPC
facility. RS thanks IUCAA for hosting his visits. We thank the WMAP
team for producing excellent quality CMB maps and making them publicly
available. We thank Amir Hajian, Subharthi Ray and Sanjit Mitra in
IUCAA for helpful discussions. We are grateful to Lyman Page, Olivier
Dore, Francois Bouchet, Simon Prunet, Charles Lawrence and an
anonymous referee for thoughtful comments and suggestions on this
work.  PJ and RS thank Sudeep Das for collaborating during the initial
stages of this project.

\end{document}